# L-Cysteine Polymorph Coatings for THz Sensing Metasurfaces

M. Zhezhu[1], A. Vasil'ev[1], H. Parsamyan[2], T. Abrahamyan[2], G. Baghdasaryan[1], S. Gyozalyan[1], D. A. Ghazaryan[3,4], H. Gharagulyan[1,2,*]

[1]*A.B. Nalbandyan Institute of Chemical Physics NAS RA, 5/2 P. Sevak str., Yerevan 0014, Armenia*
[2]*Institute of Physics, Yerevan State University, 1 A. Manoogian, Yerevan 0025, Armenia*
[3]*Moscow Center for Advanced Studies, Kulakova str. 20, Moscow, 123592, Russia*
[4]*Laboratory of Advanced Functional Materials, Yerevan State University, Yerevan 0025, Armenia*

**Author to whom correspondence should be addressed:* hermineghararagulyan@ysu.am

**The electromagnetic response of metasurfaces can be intentionally engineered by carefully designing their unit cells. Narrowband metasurfaces, characterized by high quality factors, can serve as an efficient platform for biosensing in optical to THz regimes, to explore new structural forms of biomolecules, further elevating the capabilities of THz sensing technologies. In this study, an all-metallic metasurface featuring structural asymmetry is proposed to analyze L-Cysteine with orthorhombic and monoclinic crystallographic structures. The influence of the concentration of L-Cysteine in the monoclinic phase on the THz response of the metasurface was performed using a drop-casting method. Our findings reveal a noticeable frequency shift as analyte concentration raises, underscoring the potential for high sensitivity in detecting biomolecules. Thus, this research demonstrates that asymmetric THz metasurfaces offer rather promising detection capabilities at lower THz frequencies between two crystallographic structures of L-Cysteine, with a Q-factor of 37, and a sensitivity of 1.430 THz ml/g. We believe that our results can be of great value to the development of biosensors for improved material characterization at low frequencies.**

**Abbreviations:** Cys: L-Cysteine; MS: Metasurface; THz: terahertz; THz-TDS: terahertz time-domain spectroscopy; MUT: material under test.

## Introduction

Metasurfaces (MSs) have attracted considerable interest in recent years due to their ability to manipulate the main characteristics of electromagnetic waves at sub-wavelength scales [1-3], such as reflection/transmission amplitude [4], phase [5], and polarization [6]. These properties mainly stem from carefully engineered unit cells comprising sub-wavelength metal and/or dielectric resonators [7-8] and resulting in a specific collective electromagnetic response. To date, a tremendous number of MSs have been suggested [9, 10] operating across the electromagnetic spectrum from visible light to microwaves [11-13]. Among these, MS contributed enormously to filling the so-called THz gap by enabling the development of various high-performance devices, such as filters, absorbers, waveplates, *etc.* [3, 14].

Many biological samples, including those found in the human body, have unique THz responses due to matching between the vibrational and rotational energy levels of biomolecules and photons [15] making THz technologies important for developing functional biosensors. To realize an efficient MS-based sensor, a narrowband (high *Q*-factor) resonant response of an MS needs to be engineered. The appearance of narrowband resonances in the transmission or reflection spectrum of an MS relies on several phenomena, including Fano resonances [16, 17], quasi-bound states in the continuum [18, 19], and surface lattice resonance [20]. In sensing applications, narrowband MSs have been proposed for protein, water-methanol mixtures, and DNA sensing [17, 29, 30].

Characterization of samples in the THz spectrum can be carried out by utilizing THz time-domain spectroscopy (THz-TDS) providing an efficient platform for analysis of electrical and optical properties of materials, charge carrier dynamics, and molecular fingerprints [23, 24]. In particular, it is widely used as an effective tool for amino acid detection [25] and classification [26, 27], as well as for their quantitative measurements [28] and temperature dependence studies [29, 30]. From this standpoint, exploiting MS sensors paves the way to improvement of the detection sensitivity of biomolecules and cells [31, 32]. This technology mostly solves the issues of long detection period, high loss, and low sensitivity concerning traditional amino acid detection methods, such as gas and liquid chromatography. Notably, the detection of some amino acids, such as glycine, L-arginine, and L-threonine, is discussed in [33].

Among all amino acids, L-Cysteine (Cys) is an ideal target for THz biosensing, pharmaceutical studies, and advanced material characterization, because of its distinct vibrational modes, thiol-based interactions, sensitivity to environmental changes, and role in biological processes in general [30, 34, 35]. When Cys is exposed to irradiation, it exhibits interesting behavior, such as reduction, amorphization, crosslinking, polymerization, radical formation, functional group modification, *etc*... [36, 37]. Another interesting case of Cys is its dimerized form, which has a wide range of applications [38]. Besides, Cys exhibits multiple polymorphic structures at low temperatures [39] and high pressures [40]. Notably, under ambient conditions, two crystalline polymorphs of Cys can be obtained: an orthorhombic and monoclinic phase [41].

In this work, a sensing platform of a brass MS with structural asymmetries was fabricated and the change in its THz resonance properties under the influence of Cys coating were studied *via* THz-TDS. The analysis showed that the performance of this structure is mainly dependent on the crystallographic structure of Cys, which can be either orthorhombic or monoclinic. A comprehensive characterization of orthorhombic and monoclinic phases of Cys through XRD and Raman spectroscopy were also performed. In addition, the sensing performance of our platform, such as $Q$-factor, sensitivity, and detection precision were estimated for the optimal Cys concentration.

## 1. Materials and Methods

**Materials.** The brass foil, L-Cysteine > 99 % (CAS No.: 52-90-4), solvents, and all the chemicals used in this work were purchased from Sigma-Aldrich Chemical Co.

### 1.1. Characterization

Crystallographic data for Cys polymorphs were obtained through XRD analysis (Rigaku MiniFlex instrument with CuK$\alpha$ – radiation) and its molecular conformations were analyzed *via* Raman spectroscopy (LabRAM HR Evolution, HORIBA) using a 633 nm HeNe red laser with power of 0.5 mW.

### 1.2. Spectral Response in THz range

The spectral response of the material under test (MUT) in the THz range was carried out by the all-fiber-coupled THz time-domain spectrometer (TERA K15, Menlo Systems GmbH). The experimental setup of the THz-TDS is presented in Fig. 1.

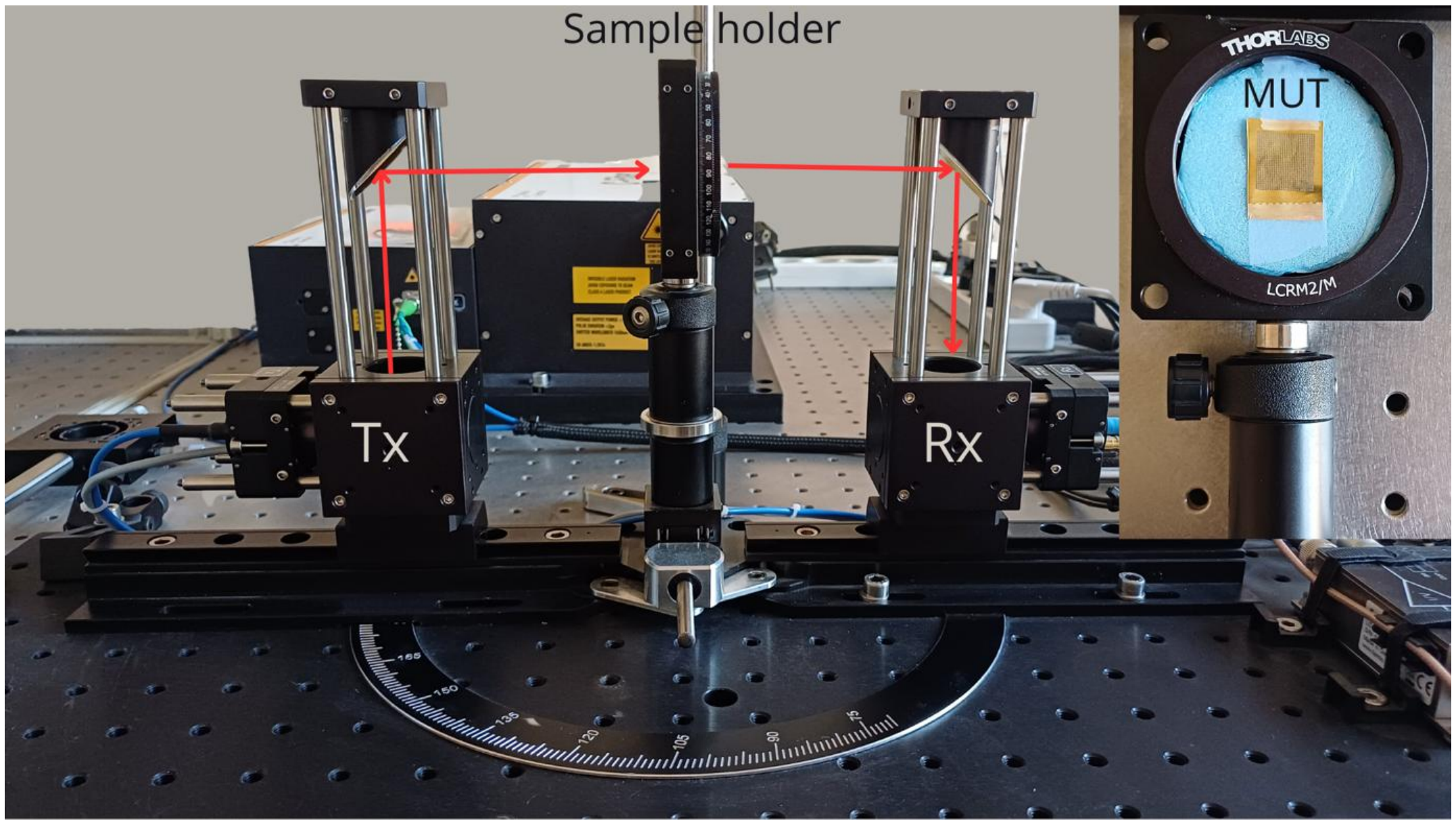

Fig. 1. Experimental setup of the THz-TDS system. Tx and Rx denote the transmitter and receiver antennas, respectively. The inset shows the sample holder with the material under test (MUT).

Here, THz signal was generated by the 1560 nm femtosecond laser pulse (100 MHz, 90 fs) using a dipole photoconductive antenna. A linearly polarized THz wave was then focused onto the sample using parabolic mirrors. The THz beam spot size at the focal plane was approximately 3 mm. The scanning range in our experiments was 280 ps, which is corresponding to a spectral resolution of about 1.83 GHz. The frequency-domain spectrum was eventually derived using a Fourier transform of the detected transmitted signal. The transmitted signal through the air was used for normalization. The MUT was attached to a foam polymer (Polypropylene) and placed at the center of the rotatable sample holder between the antennas. Note, that the polymer is highly transparent in the THz range.

### 1.3. Metasurface Geometry, Modelling and Fabrication

The free-standing all-metallic MS supporting a Fano-like high quality ($Q$) factor resonance was adopted to detect the response of low-concentration Cys in THz range. The suggested MS was based on the complementary design and was composed of two rectangular apertures with different lengths engraved in a thin metallic sheet. First, the structure was numerically analyzed in the finite element method-based COMSOL Multiphysics commercial environment. For this, it was modelled by applying periodic boundary conditions to the interfaces of the single unit-cell parallel to the propagation direction. The system was sandwiched by perfectly matched layers to eliminate secondary-reflected waves. Further details of the simulation setup are provided in Supplementary Materials (Fig. S1). The system was sandwiched by perfectly matched layers to eliminate secondary-reflected waves. The schematics of the unit cell of the MS, as well as its optical image of an array of 6 × 6 elements in the fabricated sample (the details will be provided further), are shown in Fig. 2 (a). $L_1$ and $L_2$ denote the lengths of apertures, $w$ their width, $g$ is the center-to-center distance between the apertures, and $P$ is the unit-cell periodicity in both $x$- and $y$-directions. The MS was optimized by changing the distance between apertures and the periodicity to achieve a higher $Q$-factor for efficient sensing while also considering the laser engraving system parameters. The optimized parameters obtanied were: $L_1$ = 340 μm, $L_2$ = 380 μm, $w$ = 40 μm, $g$ = 80 μm, and $P$ = 450 μm. The thickness of the metallic sheet was 50 μm. During the simulations, the conductivity of the metal was set at $10^7$ S/m, which is typical for THz response in conventional metals. The MS was fabricated using laser engraving of 50 μm thick brass sheet considering the optimized parameters obtained during the numerical analysis. Particularly, the periodic apertures in the brass sheet were created by employing a high-power nanosecond Ti:Sapphire (at 1064 nm) pulsed laser (CCI Laser, Shandong CCI Co. LTD). At the first step, the metasurface pattern was created in a dedicated software considering the laser spot diameter. Next, the laser parameters are set up including speed, power, and repetition rate for the chosen metal. The fabricated MS consists of a total of 30 × 30 elements in the array of identical unit-cells with a total area of 13.5 × 13.5 $mm^2$. The experimentally and numerically obtained normalized transmittance spectra of are shown in Fig. 2 (b) by black and red lines, respectively. Here, one sees a strong correspondence between these results. In particular, a deep in the transmittance spectrum is observed at about 0.386 THz.

The operation principle of an MS composed of asymmetric rectangular apertures can be described by a simple model of coupled oscillators using the second-order differential equations as follows [42, 43]:

$$\ddot{x}_1(t) + \gamma_1\dot{x}_1(t) + \omega_{01}^2 x_1(t) - \kappa\dot{x}_2(t) = \alpha E(t), \quad (1)$$

$$\ddot{x}_2(t) + \gamma_2\dot{x}_2(t) + \omega_{02}^2 x_2(t) - \kappa\dot{x}_1(t) = 0, \quad (2)$$

where $x_i$ is the oscillation amplitude, $\gamma_i$ is the damping factor, $\omega_{0i}$ is the resonance frequency, $\kappa$ is the coupling coefficient, and '*' denotes the complex conjugate ($i$ = 1, 2).

The insets in Fig. 2 (b) show the simulated distributions of the normalized electric (left) and magnetic (right) field magnitudes at the resonance frequency of 0.386 THz. A strong localization of the electromagnetic field is observed in both apertures due to the coupling between them. The $Q$-factor of the MS can be evaluated according to the expression: $Q \approx f_0/\delta f$, with $f_0$ and $\delta f$ being the resonance frequency and full width at half maximum, respectively. The experimentally obtained $Q$-factor of MS was approximately 37, whereas that of the numerical model was 485. The difference between the experimentally obtained Q-factor and the numerical one lies in the fact that, in the numerical model, losses are only attributed to the finite conductivity of the metal, whereas the fabricated sample exhibits additional loss channels due to fabrication imperfections and non-uniform sizes of the corresponding apertures in the unit-cells.

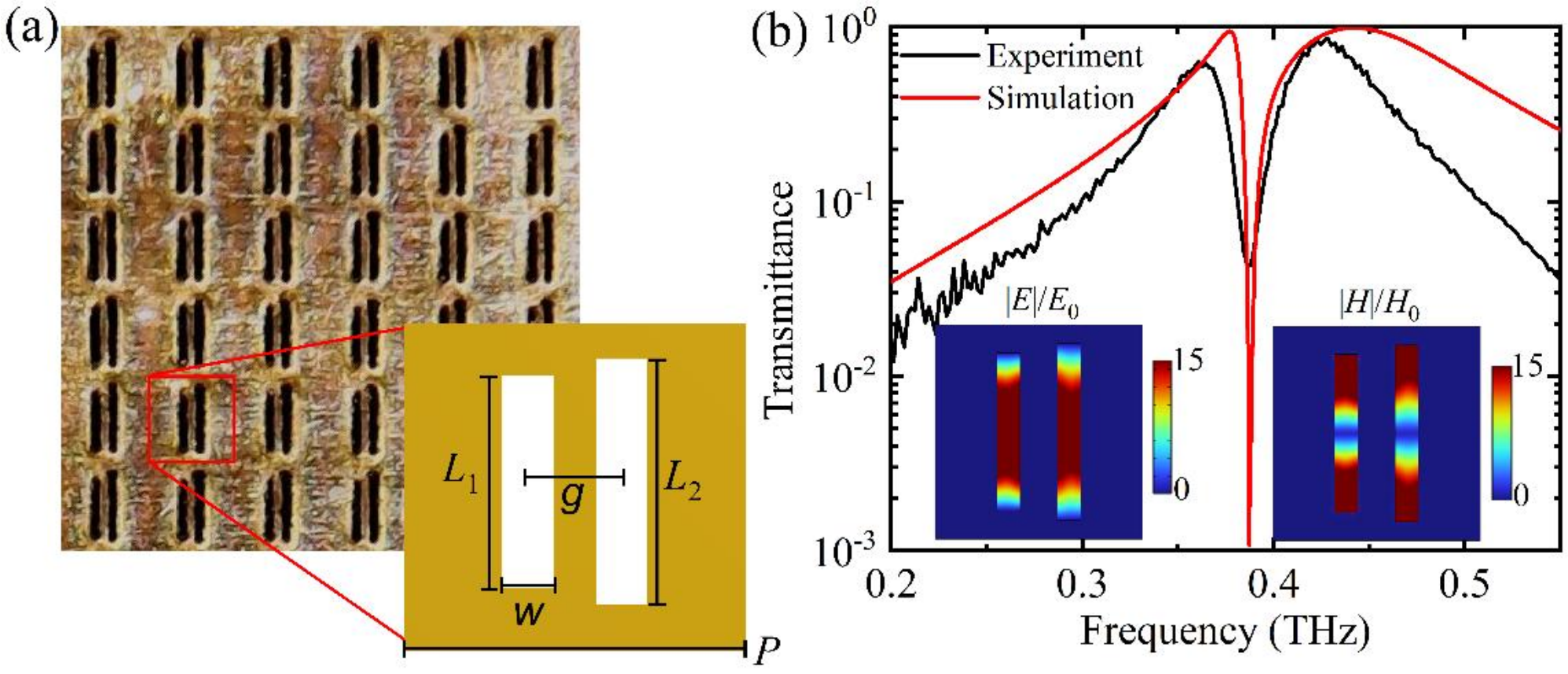

Fig.2. Fabricated brass MS and its spectral response in THz range. (a) Optical image of the designed MS and geometric parameters of its structural unit cell: $L_1$ = 340 μm, $L_2$ = 380 μm, $w$ = 40 μm, $g$ = 80 μm, $P$ = 450 μm. (b) Experimental (black) and simulated (red) transmittance spectra of the MS with resonance peaks at 0.386 and 0.387 THz, respectively.

The case of oblique incidence was also considered, and a detailed analysis of the simulation was carried out for both TM and TE polarizations. As shown in Fig. S2, TM waves exhibit angle- and polarization-dependent resonances (a), while for TE waves, resonances are absent (b). Notably, as the incident angle increases for TM polarization, the resonance peak shifts toward lower frequencies, and the depth of the dip decreases. One notes that the TE-polarized THz waves do not interact with the metasurface, whereas the TM-polarized waves demonstrate resonant interaction. This is attributed to the complementary nature of the considered structure, where the Babinet principle can be applied [44]. In this case, the apertures will interact with the electromagnetic waves whose magnetic field polarization is aligned with the long sides of the apertures, which function as magnetic dipole antennas [45]. Table S1 depicts the values of the observed phenomenon.

### 1.4. Drop-Casting L-Cysteine

First, the surface of 50 μm thick planar brass was thoroughly cleaned with alcohol and then dried. For the THz-TDS experiments, the Cys-DI water solution with different concentrations, namely, 0.03 M; 0.05 M; 0.10 M; and 0.15 M was uniformly coated on the MS by the standard drop casting method [33, 46]. In particular, 50 µl of Cys solution was dropwised onto MS and dried in ambient conditions. The measurements were completed by gradually increasing the concentration of Cys.

## 2. Results and Discussion

### 2.1. Structural Analysis of Drop-Casted L-Cysteine

Here, we discuss how Cys can adopt different polymorphs, specifically, an orthorhombic or a monoclinic phases depending on the preparation conditions [41]. These polymorphic structures influence its THz range biosensing properties by exhibiting variations in crystal packing, hydrogen bonding, and dipole interactions, ultimately improving molecular detection and biosensing applications. Structurally, orthorhombic Cys contains one molecule per asymmetric unit and represents a room-temperature disordered structure, whereas monoclinic Cys comprises two independent molecules per unit cell [39]. In the orthorhombic phase, the molecule adopts a gauche ($g^+$) conformation. In contrast, the monoclinic phase exhibits two distinct molecular conformations differing in their N–C–C–S dihedral angles: one molecule retains the gauche ($g^+$) conformation with an angle of 74.4°, while the other adopts a trans (t) configuration with an angle of −170.2°. Both polymorphs exhibit S–H···S and S–H···O hydrogen bonding, yet in the monoclinic phase, these interactions are molecule-specific—one molecule participates in S–H···O hydrogen bonding, whereas the other engages in S–H···S interactions [39]. Notably, the formation of monoclinic phase of Cys has been earlier reported following the neutralization of Cys hydrochloride with sodium hydroxide in its turn followed by the crystallization from hot water [47], slow cooling of a warm saturated aqueous solution [48], or slow evaporation from an aqueous/ethanol solution [49].

Raman spectroscopy is a sensitive tool for the monitoring changes in molecular conformations revealing that the differences observed in collected spectra of Cys powder (original from factory) and Cys drop-casted samples are due to the the formation of orthorhombic and monoclinic phases of Cys, respectively [41, 50, 51]. Fig. 3 (a) and (b) presents Raman spectra of the initial powder and drop-casted Cys. As can be seen from the spectra, those exhibit significant differences. Beyond 2500 $cm^{-1}$, the powder displays two C–H stretching bands at 2968 $cm^{-1}$ and 3001 $cm^{-1}$, while the drop-casted sample shows three distinct bands at 2930 $cm^{-1}$, 2955 $cm^{-1}$, and 2986 $cm^{-1}$, respectively. Similarly, S–H stretching vibrations differ: the powder features a broad band centered at 2552 $cm^{-1}$, whereas the drop-casted sample reveals a sharper and more intense peak at 2546 $cm^{-1}$ highlighting the differences in hydrogen bonding environments. Below 1500 $cm^{-1}$, molecular deformation vibrations also exhibit notable discrepancies. In the drop-casted sample, three bands corresponding to C–S stretching vibrations are observed at 622 $cm^{-1}$, 644 $cm^{-1}$, and 682 $cm^{-1}$, compared to two bands at 644 $cm^{-1}$ and 696 $cm^{-1}$ in the powder form. For drop-casted Cys, peaks corresponding to C–C–C rocking, C–C–C bending, and C–O–O wagging vibrations are absent compared to the powder. Conversely, new peaks emerge, such as those associated with C–O–O rocking (518 $cm^{-1}$), $CH_2$ rocking (744 $cm^{-1}$), and symmetric C–O–O stretching (1365 $cm^{-1}$). Overall, the spectral changes, including the appearance or disappearance of specific peaks and variations in peak intensities, reflect differences in molecular conformations and bonding environments. The detailed vibration assignments of both crystallographic phases of Cys are tabulated in Table S2 (see in Supplementary Materials).

To investigate changes in the molecular conformations of Cys, an XRD analysis was also done for the initial powder and drop-casted samples. It was revealed that Cys powder is crystallized in orthorhombic symmetry (predominant reflection of lateral planes) and for the drop-casted sample, a monoclinic orientation was observed (see Fig. 3 (c) and (d), respectively) [47, 52]. Notably, the low-symmetry monoclinic phase is more preferable due to its lowest crystallization energy. The anisotropic nature of Cys monoclinic platelets facilitates crystallization in a highly oriented manner along the *ab* crystallographic plane. As a result, we observed strong XRD peaks from the *c*-plane confirming the textured nature of the material.

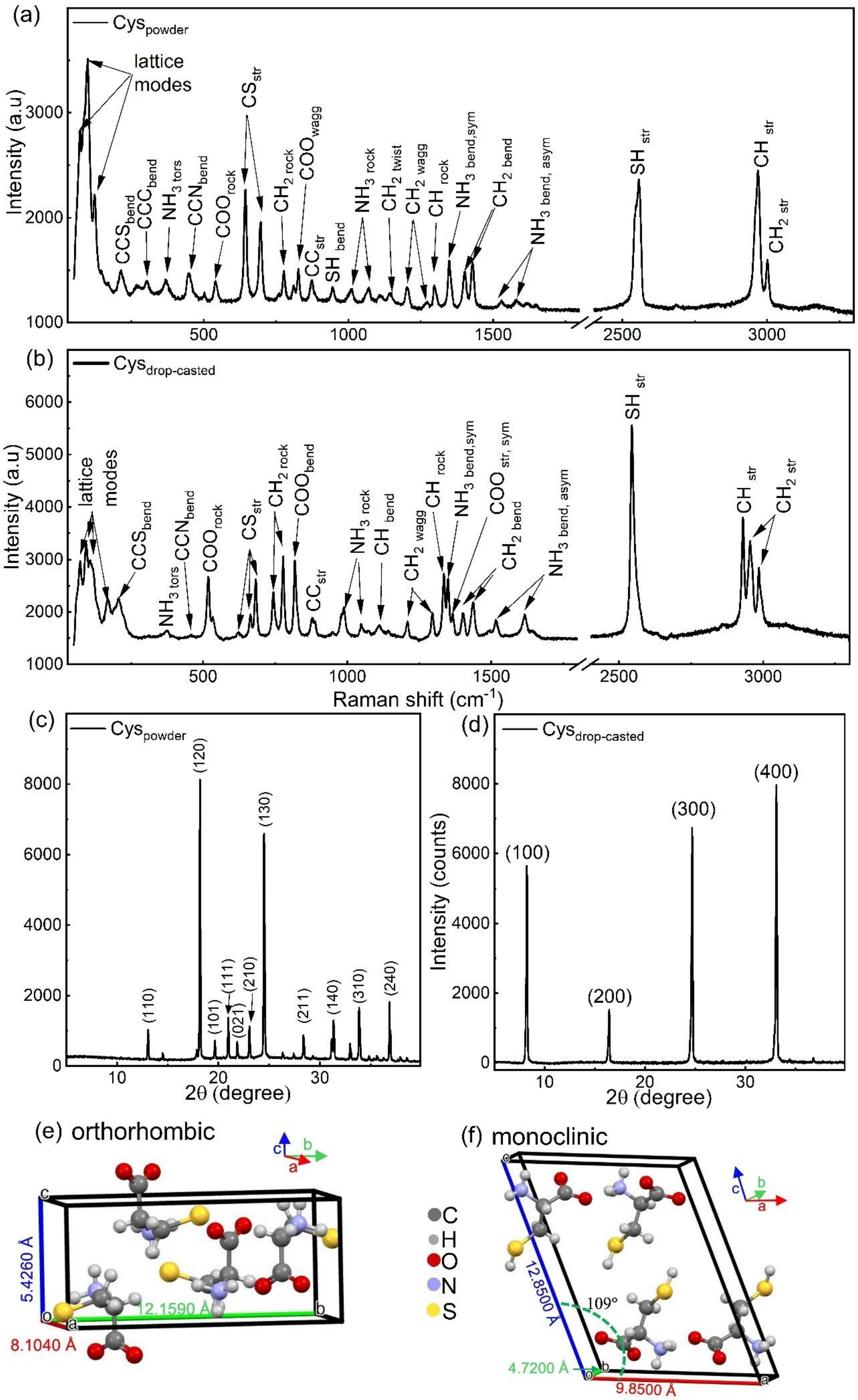

Fig. 3. Structural characterization of Cys powder and drop-casted samples: Raman spectra of (a) Cys powder and (b) drop-casted Cys measured with a 600 grating, an acquisition time of 20 s. XRD spectra of (c) Cys powder and (d) drop-casted Cys. The measurements were performed at temperature of 25°C. An illustration of (e) orthorhombic and (f) monoclinic structures of Cys.

Fig. 3 (e) and (f) shows the two structural phases of Cys: orthorhombic and monoclinic, respectively, implemented from CCDC (Cambridge Crystallographic Data Centre) database. The orthorhombic structure of Cys crystallizes with Z = 4 in $P2_12_12_1$ space group (PDF Card No.: 00-032-1636). Whether the monoclinic phase of Cys crystallizes with Z = 4 in $P2_1$ space group (PDF Card No.: 00-024-1928). The corresponding lattice parameters were calculated using the Smart Studio 2 software from the experimental data employing Rietveld's method. For orthorhombic Cys, lattice parameters were $a$ = 8.1040 Å, $b$ = 12.1590 Å, $c$ = 5.4260 Å, $\alpha$, $\beta$, $\gamma$ = 90°, and for monoclinic Cys, lattice parameters were $a$ = 11.5027 Å, $b$ = 5.2893 Å, $c$ = 9.4960 Å, $\alpha$, $\gamma$ = 90°, $\beta$ = 109°, respectively [35, 53].

## 2.2. THz-TDS Analysis of L-Cysteine Polymorphs on the Metasurface

The THz response of MUT was experimentally analyzed by using a THz-TDS system as depicted in Fig. 1. First, we investigated the THz responses of drop-casted Cys on a polyethylene terephthalate (PET) substrate with the lowest and highest concentrations. The thickness of THz-transparent PET substrates was 0.25 mm. The results are depicted in Fig. 4 (a), where the THz spectra of the samples were normalized to the transmittance of PET substrate. The transmittance of the bare PET and its temporal profile are shown in Fig. S3 (see Supplementary Materials).

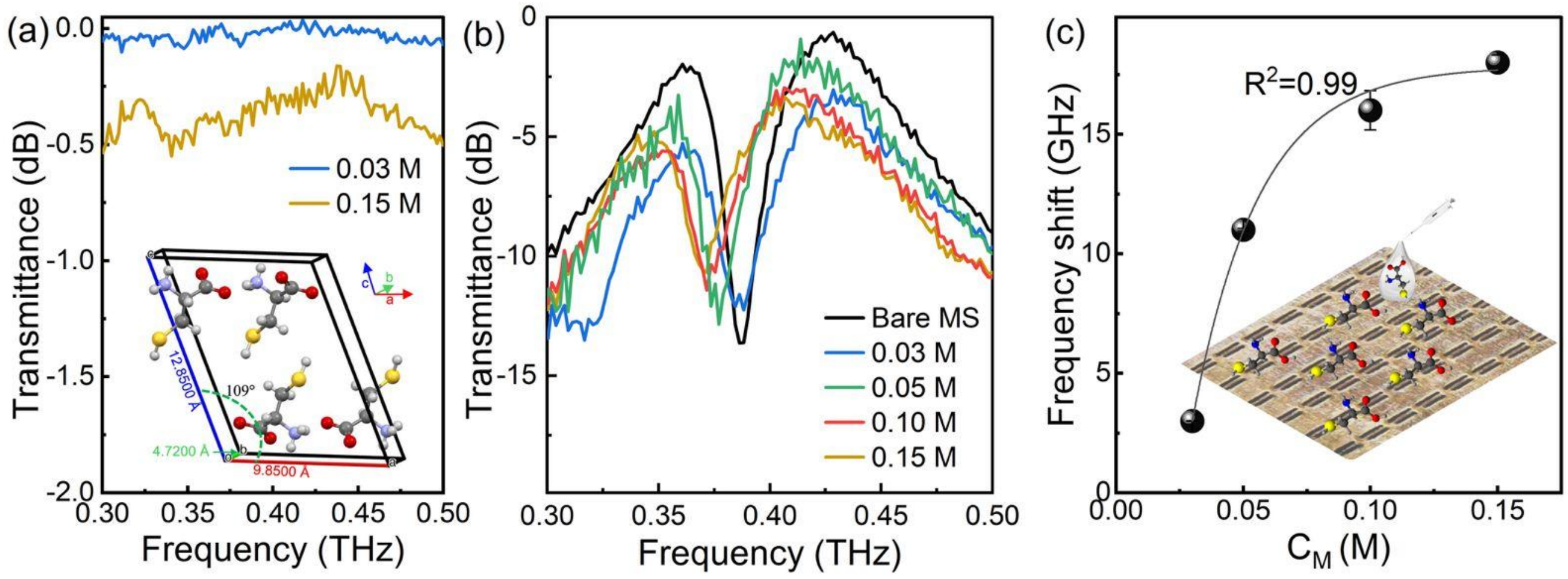

Fig. 4. The THz spectrum of the drop-casted Cys and quantitative sensing features: (a) Transmission spectra of monoclinic phase of Cys on PET substrate. The inset shows the structure of monoclinic Cys. (b) The transmission spectra of the MS sensor for different concentrations of monoclinic Cys. (c) The resonance frequency shift as a function of Cys concentration. A nonlinear correlation is modeled using a nonlinear fit $y = -46.9e^{(-38.4x)} + 17.8$ $y = -46.9e^{(-38.4x)} + 17.8$.

The inset in Fig. 4 (a) shows the monoclinic crystallographic structure of drop-casted Cys. It's worth mentioning that at low frequencies and the lowest concentration, Cys is not detectable due to its weak signal response. To effectively detect the low-concentration Cys at low frequencies, we used a complementary MS sensor as described earlier. A sensor performance is typically evaluated using three key parameters: sensitivity ($S_f$), (determined by $S_f = \frac{\Delta f}{\Delta C}$, where $\Delta f$ is the frequency shift of the resonance peak, and $\Delta C$ represents the corresponding change in sample concentration [54], $Q$-factor described above, and detection precision, estimated as $M_c = \frac{\Delta}{S_f \cdot Q}$, where $\Delta$ represents the frequency resolution of the system [55]. Since the response of the suggested MS sensor is conditioned by the interaction of THz signal with rectangular apertures and their mutual coupling, any changes in the aperture refractive index will eventually result in a resonance shift. Cys, through its thiol side chain, can form chemical bonds with the brass surface serving as a molecular bridge between multiple biomolecules and metal surfaces [56]. The experimentally measured transmission spectra of the MS sensor without (bare) and with a drop-casted Cys of varying concentrations (0.03, 0.05, 0.1 and 0.15 M) are presented and compared in Fig. 4 (b). Here, one observes a noticeable shift of the transmission deep towards lower frequencies as the Cys concentration increases. Table S3 in Supplementary Materials sums up the resonance frequency for each Cys concentration considered, as well as the resonance frequency of the MS with Cys powder. On the other hand, as shown in Fig. 4(a), the 0.03 M Cys layer is practically transparent to THz waves within the spectrum of interest, whereas the transmittance of the 0.15 M Cys layer is approximately -0.5 dB, corresponding to a linear transmittance of about 0.89. Fig. 4(b) further reveals that increasing the Cys concentration results in a shift of the resonance dip towards lower frequencies. This indicates that the refractive index of Cys increases with concentration. Therefore, the observed change in transmittance of the drop-casted Cys layer can primarily be attributed to increased reflection.

The alignment between the resonance peak of MS sensor and the fingerprint peak of the sample is one of the primary factors influencing the sensitivity [33]. Based on CASTEP-calculated transition energies, Cys exhibits energy transitions at 13 $cm^{-1}$ in the orthorhombic phase and 37 $cm^{-1}$ in the monoclinic phase, which corresponds to the translation mode [41]. Importantly, the orthorhombic phase suggests weaker molecular interactions and less dense arrangement, while the monoclinic one - stronger molecular interactions, *i.e.*, tighter packing. These differences arising due to crystal packing forces, such as hydrogen bonding and van der Waals interactions, have interesting applications in THz spectroscopy for studying the molecular interactions, phase transitions, and material properties [57]. On the other hand, MS features a resonance at 0.386 THz corresponding to the transitions at ≈ 12.9 $cm^{-1}$, which can result in the resonance broadening. The dependence of the resonance frequency on Cys concentration is plotted in Fig. 4 (c). As can be seen from the figure, the frequency shift becomes more pronounced with increment of the concentration, but the rate of change gradually slows down. The observed saturation results from the greater accumulation of the dried sample at higher concentrations, which reduces its impact on the overall equivalent capacitance of MS ultimately leading to a reduced frequency shift [55]. A nonlinear relationship between the frequency shift and analyte concentration has also been reported in [46, 58], an effect commonly observed in biological experiments.

To evaluate the sensing performance of our system, we estimated the sensitivity and the detection precision. The optimal sensing parameters were achieved for the detection of 0.05 M Cys yielding to $S_f \approx 1.430$ THz ml/g. With experimental spectral resolution of 1.83 GHz, the detection precision for this experimental measurement was estimated as $3.5 \cdot 10^{-5}$ g/ml. Notably, our results align with previous studies of THz sensing of amino acids and other biomolecules. For instance, in polarization-dependent THz phase shift measurements of arginine at a concentration of 40 mg/ml, the detection precision ranged from $2.5 \cdot 10^{-5}$ to $7.9 \cdot 10^{-5}$ g/ml depending on the *x*- and *y*-polarization directions [55]. Similarly, L-valine exhibited a sensitivity of 9.98 GHz/(mmol/L) with an experimental spectral resolution of 9 GHz [59]. Beyond amino acids, in biomolecular sensing, bovine serum albumin (BSA) detection has demonstrated a sensitivity of 72.81 GHz/(ng/$mm^2$) with a detection limit of 0.035 mg/mL [8], while another study reported a sensitivity of 95 GHz/(mmol/L) and a detection resolution of 17.7 µmol/L [60]. For the case of vitamin detection, the reported limits for vitamins *C* and *B9* were 158.82 ng/µL and 353.57 ng/µL, respectively [46]. These further supports the relevance of our study in the context of THz-based sensing applications.

Similar measurements were also performed for Cys powder; however, with a limitation considering that it can't be prepared in varying concentrations and detected in small amounts. The transmission spectrum of the Cys powder on PET substrate (see Fig. S4 (a) in Supplementary Materials), as well as the comparison of MS response without and with Cys powder are shown in Fig. S4 (b).

## Conclusion

In conclusion, we studied the orthorhombic (powder) and monoclinic (drop-casted) Cys using XRD, Raman and THz time-domain spectroscopies. For the latter, we adopted an all-metallic MS sensor based on asymmetric rectangular apertures supporting a high-$Q$-factor Fano resonance using a drop-casted Cys of different molarities to detect the resonance shift due to the variation of molar concentration. The MS was optimized using numerical methods to tune the resonance to the desired frequency and achieve a high $Q$-factor. The experimental value of the latter was estimated to be 37. The sensing performance of our system, including the $Q$-factor, sensitivity, and detection precision, were evaluated based on the experimental results. The analysis suggests that increasing the concentration in drop-casted Cys on MS results in a shift of the resonance frequency from 0.386 (bare MS) to 0.37 for 0.05 M solution with an estimated sensitivity of 1.43 THz ml/g and detection precision of $3.5 \cdot 10^{-5}$ g/ml. This study demonstrates that the resonance shift is primarily driven by refractive index changes due to concentration variations and crystallographic structure alterations caused by deposition of Cys and highlighting the effective detection capabilities of asymmetric MS at low frequencies. In addition, unlike MSs designed for disposable sensing, our all-metallic structure-based sensor offer reusability, as it can be easily cleaned with common organic solvents after each use.

# Supplementary Materials for

# *L-Cysteine Polymorph Coatings for THz Sensing Metasurfaces*

M. Zhezhu[1], A. Vasil’ev[1], H. Parsamyan[2], T. Abrahamyan[2], G. Baghdasaryan[1], S. Gyozalyan[1], D. A. Ghazaryan[3,4], H. Gharagulyan[1,2,*]

[1]A.B. Nalbandyan Institute of Chemical Physics NAS RA, 5/2 P. Sevak str., Yerevan 0014, Armenia
[2]Institute of Physics, Yerevan State University, 1 A. Manoogian, Yerevan 0025, Armenia
[3]Moscow Center for Advanced Studies, Kulakova str. 20, Moscow, 123592, Russia
[4]Laboratory of Advanced Functional Materials, Yerevan State University, Yerevan 0025, Armenia

*Author to whom correspondence should be addressed: herminegharagulyan@ysu.am

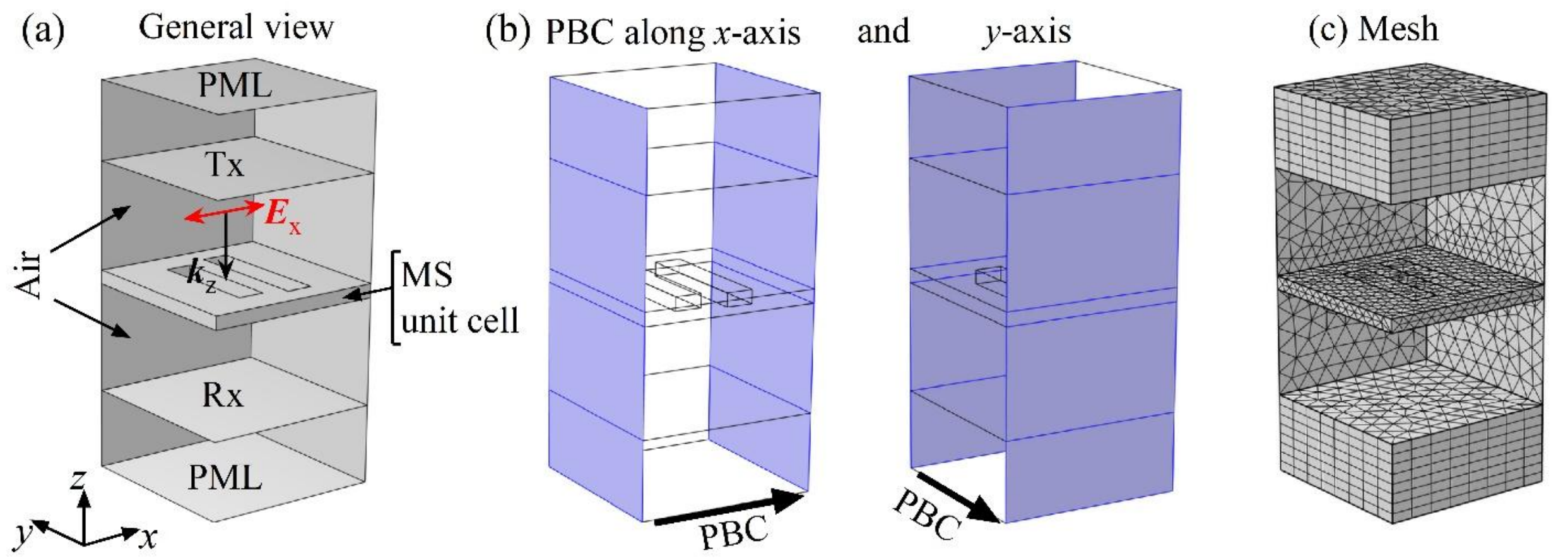

Fig. S1. Simulation setup of the metasurface. (a) General view of the simulation domain. The system is sandwiched by PMLs to effectively die out transmitted and reflected waves after passing through the ports. Tx and Rx stand for the excitation and detection ports. (b) Two sets of periodic boundary conditions along x- and y-axes. PBCs ensure the electric and magnetic fields to be equal at the corresponding opposite boundaries. (c) Discretization of the simulation domain by rectangular prisms (PMLs) and tetrahedral elements (other domains).

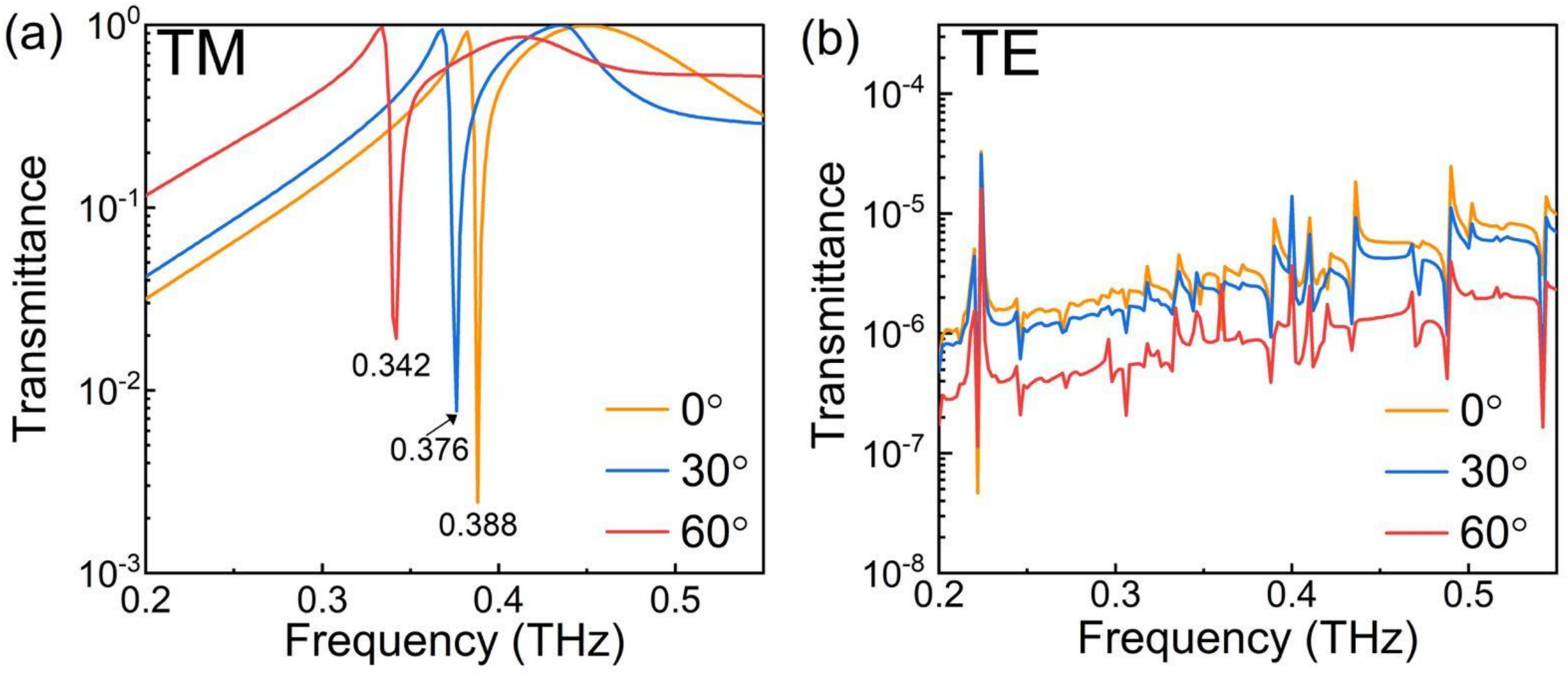

Fig. S2. Simulated transmittance spectra of the MS with resonance peaks at the different oblique incident angles for both TM and TE waves.

Table S1. Angle- and polarization-dependent resonances of the MS.

| Angle (°) | Transmittance (dB) | Frequency (THz) |
|---|---|---|
| 0 | 0.002 | 0.388 |
| 30 | 0.008 | 0.376 |
| 60 | 0.019 | 0.342 |

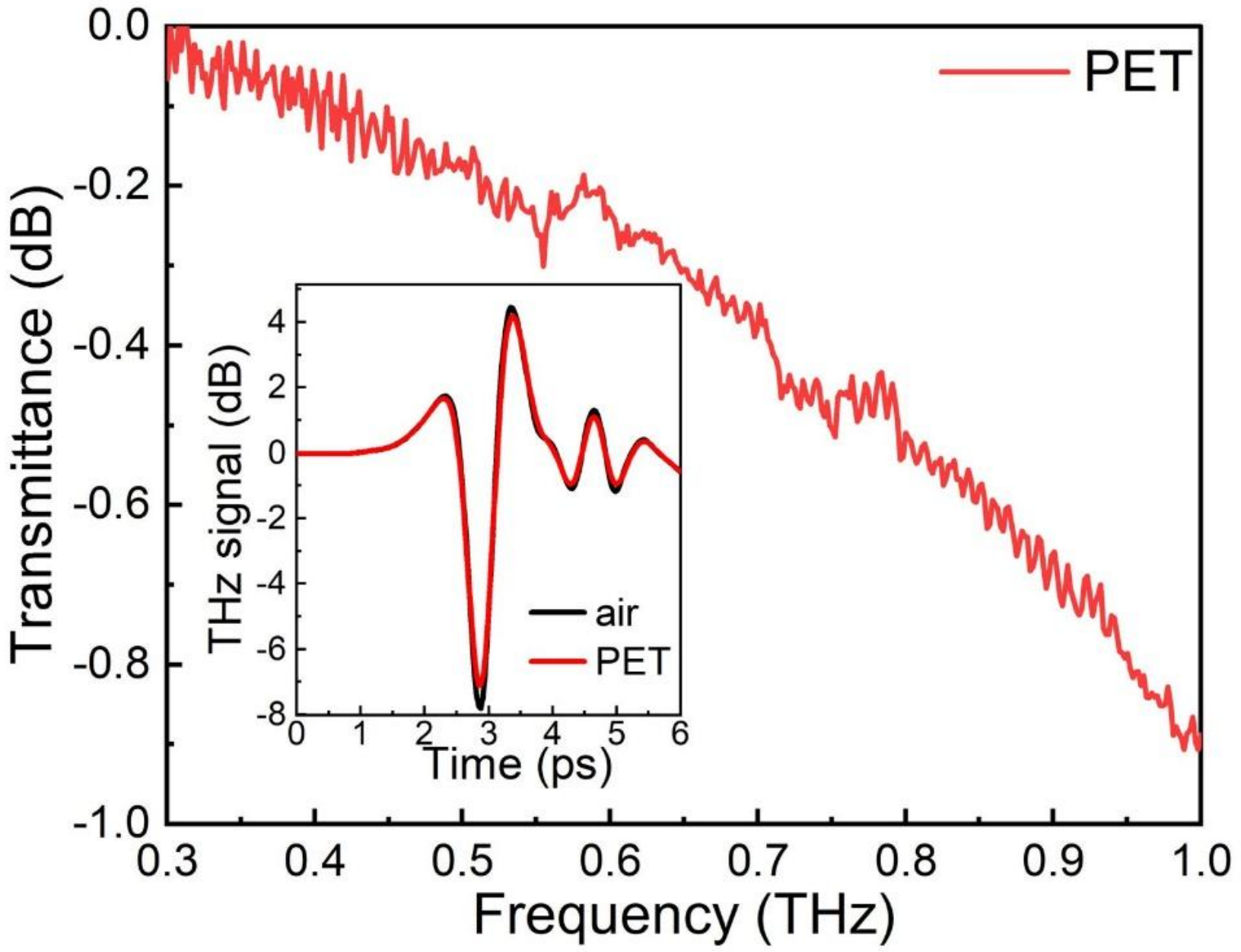

Fig.S3. Transmission spectra of the PET. Inset shows temporal profiles of a THz pulse of the PET and the reference.

Table S2. Comparison of Raman bands and their assignments for Cys in powder and drop-casted forms.

| Cys $_{powder}$ (orthorhombic) | Cys $_{drop\text{-}casted}$ (monoclinic) | Assignments |
|---|---|---|
| Raman shift ($cm^{-1}$) | | |
| 72.6 | 72.8 | Lattice modes |
| 97.7 | 94.0 | |
| 122.5 | 109.1 | |
| — | 170.2 | |
| 214.6 | 204.5 | CCS $_{bend}$ |
| 274.0 | — | CCC $_{rock}$ |
| 302.1 | — | CCC $_{bend}$ |
| 368.8 | 371.7 | $NH_3$ $_{tors}$ |
| 449.6 | 456.9 | CCN $_{bend}$ |
| — | 517.5 | COO $_{rock}$ |
| 541.2 | 558.2 | |
| — | 622.0 | CS $_{str}$ |
| 643.6 | 644.2 | |
| 696.0 | 682.3 | |
| — | 743.9 | $CH_2$ $_{rock}$ |
| 776.9 | 777.3 | |
| 811.0 | 818.9 | COO $_{bend}$ |
| 826.7 | — | COO $_{wagg}$ |
| 873.7 | 881.0 | CC $_{str}$ |
| 945.6 | 949.3 | SH $_{bend}$ |
| 1008.2 | 985.5 | $NH_3$ $_{rock}$ |
| 1068.2 | 1050.4 | |
| 1110.7 | 1110.9 | CH $_{bend}$ |
| 1143.3 | 1139.2 | $CH_2$ $_{twist}$ |
| 1203.1 | 1208.4 | $CH_2$ $_{wagg}$ |
| 1271.6 | 1295.6 | |
| 1298.7 | 1335.5 | CH $_{rock}$ |
| 1349.2 | 1349.4 | $NH_{3\ bend,\ sym}$ |
| — | 1365.3 | $COO_{str,\ sym}$ |
| 1402.1 | 1402.0 | $CH_{2\ bend}$ |
| 1428.8 | 1436.1 | |
| 1529.1 | 1516.4 | $NH_{3\ bend,\ asym}$ |
| 1581.6 | 1615.6 | |
| 2552.3 | 2546.2 | $SH_{str}$ |
| 2967.7 | 2930.3 | $CH_{str}$ |
| 3000.8 | 2955.3 | $CH_{2\ str}$ |
| — | 2985.9 | |

Table S3. The resonance peak frequency corresponded to the Cys in its drop-casted form and powder on the MS sensor.

| | $C_M$, M | Frequency, THz |
|---|---|---|
| Cys drop-casted | 0 | 0.386 |
| | 0.03 | 0.383 |
| | 0.05 | 0.375 |
| | 0.10 | 0.371 |
| | 0.15 | 0.368 |
| Cys powder | | 0.311 |

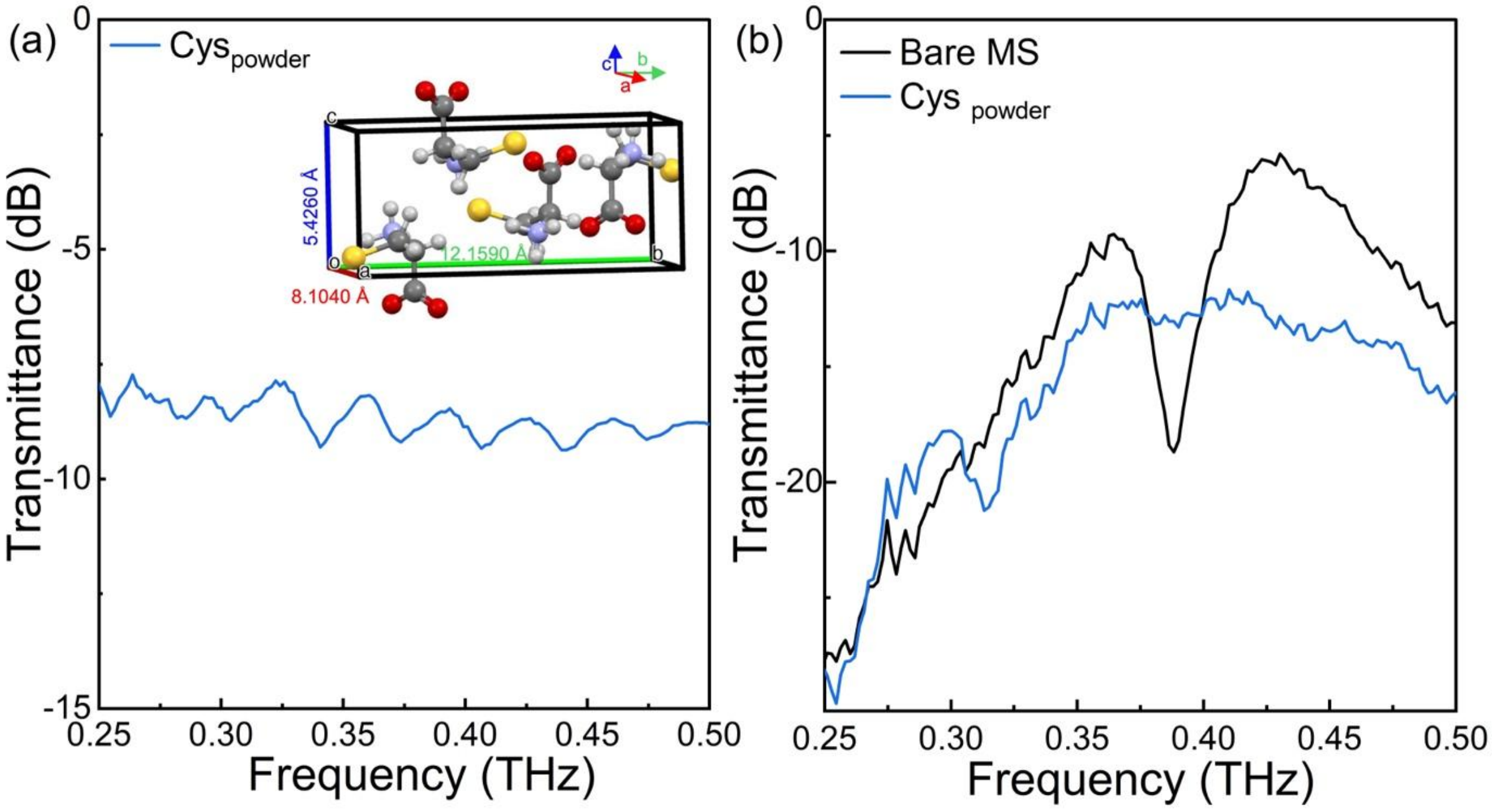

Fig. S4. (a) Transmission spectra of Cys powder on the PET and its orthorhombic crystal structure. (b) The transmission resonance spectra of the Cys powder coated MS.